\documentclass[useAMS,usenatbib]{mn2e}
\input psfig.tex

\title[Faint Red Galaxies in Abell 1158]{
New evidence for a linear colour--magnitude relation and a single
Schechter function for red galaxies in a nearby cluster of galaxies
down to $\bmath{M^*+8}$}
\author[Andreon et al.]{S. Andreon,$^1$\thanks{andreon@brera.mi.astro.it}, 
J.-C. Cuillandre,$^{2,3}$, E. Puddu,$^4$, Y. Mellier,$^5$\\
$^1$INAF--Osservatorio Astronomico di Brera, Milano, Italy\\
$^2$CFHT Corp., Kamuela, USA \\
$^3$Observatoire de Paris, Paris, France \\
$^4$INAF--Osservatorio Astronomico di Capodimonte, Napoli, Italy\\
$^5$CNRS UMR 7095, Institut d'Astrophysique de Paris, Paris, France\\
}
\date{Accepted ... Received ...}
\pagerange{\pageref{firstpage}--\pageref{lastpage}}
\pubyear{2005}
\begin{document}
\maketitle

\label{firstpage}

\begin{abstract}
The colour and luminosity distributions of red galaxies in the cluster 
Abell\,1185 ($z=0.0325$) were studied down to $M^*+8$ in the 
$B$, $V$ and $R$ bands. The colour--magnitude (hereafter CM) relation 
is linear without evidence for a significant bending down to absolute 
magnitudes which are seldom probed in literature ($M_R=-12.5$ mag). The 
CM relation is thin ($\pm0.04$ mag) and its thickness is quite independent 
from the magnitude.  The luminosity function of red galaxies in
Abell\,1185 is adequately described by a Schechter function, with a
characteristic magnitude and a faint end slope that also well describe the LF
of red  galaxies in other clusters. There is no passband dependency of the
LF shape other than an obvious $M^*$ shift due to the colour of the 
considered population. Finally, we conclude that, based on colours and
luminosity, red galaxies form an homogeneous population over four decades 
in stellar mass, providing a second evidence against faint red galaxies being
a recent cluster population. 
\end{abstract}


\section{Introduction}

One of the major problems facing models of galaxy formation is why cold dark 
matter models predict a larger number of low-mass galaxies than observed.
Today, extremely faint ($M \gg M^*+5$ mag) galaxies are still a poorly 
studied population because they are difficult to find and, once found, the 
measurement of their properties is not trivial since the measured properties 
are usually strongly affected by selection effects. Thus, understanding the
properties of the lowest luminosity galaxies, which are presumably also
very low mass, might shed light on some of the unsolved questions related to
the production of galaxies in low-mass halos.

The bulk of the observational work on low-luminosity galaxies is limited to 
clusters environment. The pioneering work by Visvanathan \& Sandage (1977) 
presents the colour-magnitude relation over an eight magnitude range, although 
the very large majority of the data is relevant to the brightest four magnitudes. 
Secker et al. (1997) reports that the colour-magnitude relation (also known as 
the red sequence) is linear in the Coma cluster over a seven magnitude range, 
i.e. down to $M_R=-15.0$ mag. Conselice et al. (2002) confirms that galaxies in 
the first four magnitudes of the Perseus clusters obey to the usual colour--magnitude 
relation (e.g. Bower, Lucey \& Ellis 1992), but faint candidate members of the cluster 
tend to depart from the colour of the red sequence, being bluer or redder. The scatter 
around the red sequence is small ($\sigma \approx 0.07$ mag) down to $M_R\sim-17$ mag,
but rises to $\sigma \approx 0.5$ mag at $M_R\sim-14$ mag. Instead, Evans et al. (1990) 
found that the fainter Fornax galaxies become redder, not bluer. Therefore, it is 
unclear what is the shape of the red sequence at faint magnitudes and if it is
universal or if it changes from cluster to cluster. 

In other environments, information is even scarcer. In the review article by Mateo (1998) 
on dwarf galaxies in the local group, the colour-magnitude relation includes about 30 
galaxies in the range $-22<V<-11$ mag. Blanton et al. (2005) derived luminosity and 
colour distributions of faint galaxies in the local universe for a flux and surface brightness
limited sample. However, it is unclear how much their results are affected by the fact
that they assume no environment dependence and a uniform spatial distribution of the 
studied sample, and later on, the distributions were claimed environmental-dependent 
and the sample clustered (see Andreon, Punzi \& Grado 2005 for a discussion).

\begin{figure*}
\psfig{figure=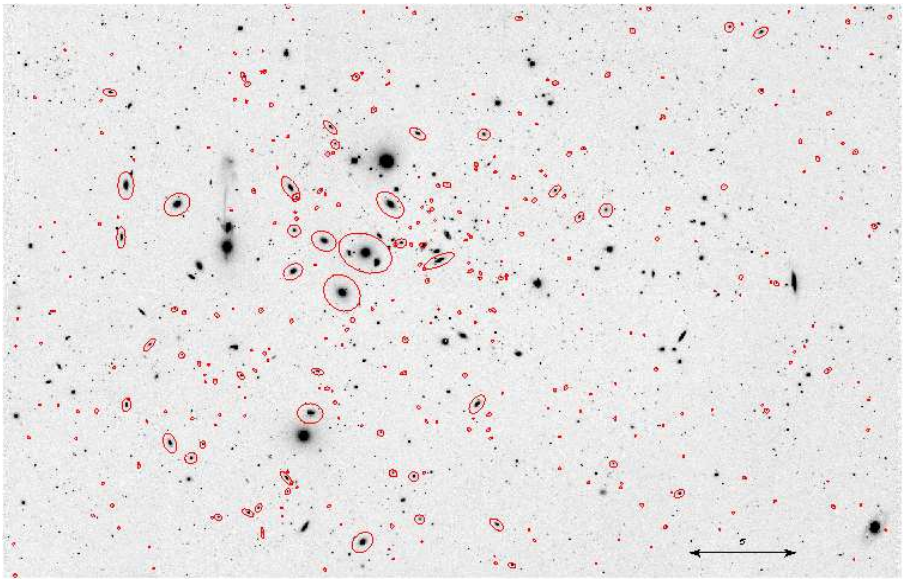,width=12truecm,clip=}
\caption[h]{(Compressed) image of the studied field in the $R$ band, with
red galaxies (members and background) marked with an ellipse. 
North is up and east is to the left. 
The field of view is $42\times28$ arcmin$^2$.} 
\end{figure*}

In this paper, we report results about a volume complete sample of
galaxies in a nearby cluster of galaxy, Abell\,1185. 
Our data are deep enough to reveal extremely faint objects, allowing us to
investigate the properties of red galaxies over a nine magnitude range.

Abell\,1185, a cluster of richness one class (Abell\,1858), has 
a redshift $cz = 9800$ km s$^{-1}$ and a velocity dispersion of 
$\sigma_v=740\pm60$ km s$^{-1}$ (Mahdavi et al. 1996), and
an X-ray luminosity of  $L_X$(0.5-3 keV) $= 1.6 \ 10^{43}$ 
$h_{50}^{-1}$ ergs s$^{-1}$ (Jones \& Forman 1984). 

For the Abell\,1185 cluster, we adopt a distance modulus of 35.7 mag 
($H_0=70$ km s$^{-1}$ Mpc$^{-1}$, $z=0.0325$).

\section{The data and data reduction}

Abell\,1185 observations and data reduction follow closely a similar analysis 
performed on the Coma clusters (Andreon \& Cuillandre 1999, AC02 hereafter), 
to which we defer for details.

$B$, $V$ and $R$ cluster observations were taken on January 31th,
2003 in photometric conditions, with the CFH12K camera (Cuillandre et al. 2000) 
at the Canada--France--Hawaii Telescope (CFHT) prime focus in photometric
conditions. CFH12K is a 12,288 $\times$ 8,192 pixel CCD mosaic camera, 
with a 42 $\times$ 28 arcmin$^2$ field of view and a pixel size of 0.206 
arcsec. Five dithered images per filter were taken, pre-reduced (overscan, 
bias, dark and flat--field) and then optimally stacked. The total exposure 
time was 1500, 1200 and 900 s in $B$, $V$ and $R$, respectively. Seeing in 
the combined images is 1.2 to 1.3 arcsec FWHM. The CFHT CCD mosaic data 
reduction package FLIPS (Magnier \& Cuillandre 2004) was used. After 
discarding areas noisier than average (gaps between CCDs, borders, regions 
near bright stars, etc.), the usable area is 0.263 deg$^2$, or 1.3 Mpc$^2$ 
at the cluster redshift. Images were  calibrated in the Bessel--Cousin--Landolt 
system through the observation of photometric standard star fields listed in Landolt 
(1992).

Figure 1 shows an highly compressed $R$ band image of Abell\,1185.

In this work, interloper galaxies are statistically removed. This requires
that the cluster and the control field images share the same photometric system:
in presence of colour mismatches, which often arise from using data coming from 
different telescopes/instruments/filters, the background contribution is usually 
assumed to be perfectly removed whereas instead it leaves some residuals, biasing 
the LF and colour measurements. Our control field has been published and described
in McCracken et al. (2003). These images were taken using the very same $B$, $V$ 
and $R$ filters, telescope and instrument used for the present study; they were reduced 
by the same software, calibrated in the same photometric system by the same standard 
fields, hence assuring a perfect homogeneity between the two data sets. The control 
field is much deeper than the cluster observations, and therefore we consider only the 
magnitude range that matches the cluster images. Even if our data are homogeneous, we 
nevertheless verified that cluster and control field images  share the same locus of stars, 
in the $V-R$ vs $B-V$ colours (e.g. as in Fig 1 of Andreon, Lovo, \& Iovino 2004), and in 
the $(V-R)-(B-V)$ vs $B-V$, in order to enhance the visibility of any mismatch. As a further 
check, we compared the density distribution of galaxies in  the $V-R$ vs $B-V$ plane, and
did not detect any colour offsets between the cluster and the control field images. 

Objects are detected using SExtractor (Bertin \& Arnouts 1996), using a
isophotal threshold  of 26.0, 25.5, 24.5 in $B$, $V$ and $R$, respectively,
corresponding to a threshold of $\approx 2 \sigma$ of the sky in the cluster
images, and more than 5 $\sigma$ in the control field images.

In this paper, we adopt SExtractor isophotal corrected magnitudes as
a proxy for ``total'' magnitudes; we do not choose Kron magnitudes because
of their flip--flopping nature described in Andreon, Punzi \& Grado (2005, APG hereafter).
Completeness is measured as the magnitude of the brightest low surface brightness galaxy of
the faintest detectable central surface brightness, as described in Garilli, Maccagni \& 
Andreon (1999). Cluster data are complete down to $B=24.0$, $V=23.8$ and $R=22.7$ mag, but
further clipped to a brighter magnitude range ($B=23.25$ or $V=22.45$ or $R=22.0$ mag, 
depending on the measured quantity), in order to be nearly 100\% complete in colour, and in
order to reduce the Eddington bias (see sec 3.5.2 in AC02 for details). Fundamentally, 
from more than 10,000-20,000 detections in the cluster line of sight, we select the brightest 
2,000 galaxies with the rationale of keeping the best data.

Stars are discarded by combining the probabilities derived by SExtractor in the three filters.
We conservatively keep a high threshold, rejecting ``sure stars" only ($class_{star}>0.95$)
in order not to exclude the compact galaxies, while leaving some residual stellar contamination in
the sample: it (as well as the contamination caused by background galaxies) is dealt with statistically later on.
Most galaxies are easily classified as extended sources in the magnitude range of our interest. Very compact 
galaxies, such as NGC\,4486B or M\,32, are however easily misclassified as stars at the Fornax (Drinkwater et al. 1999) 
and Coma (AC02) clusters distances, but they represent a minority population (Drinkwater et al. 1999, AC02). 
Finally, blends of globular clusters (GCs) that are not resolved in individual components due to seeing, and look like 
dwarf galaxies on our ground-based images, are expected, but from what is observed in the Coma cluster (AC02), 
the GCs blends start to contaminate galaxy counts at fainter magnitudes than the range probed in this work.

\section{Results}

\subsection{Exploring the colour-colour-magnitude data cube}

Let us explore the colour-colour-magnitude cube 
by observing it through different projections.

Figure 2 is a projection of the colour-colour-magnitude cube that
aligns the observer's eye along the colour--magnitude (hereafter CM) relation. 
The projection is along an axis slightly tilted with magnitude, with slopes 
$0.2/6$ and $0.1/6$ mag in $B-V$ and $V-R$, respectively (these are logically 
the slopes of the colour--magnitude relations in the two colours). Technical 
details about the way the CM slope is determined are given in the appendix.
Colours are corrected to $R=20$ mag for the slope of colour--magnitude.
The left and right panels refer to galaxies in the cluster and in the 
control field lines of sight, respectively. The number density distribution of control
field galaxies (right panel) shows a shallow gradient from bottom-right to
top-left, more evident when the whole control field data set is plotted (in
the figure, only one galaxy out of 4.7 is plotted to account for differences 
in the surveyed area and to make the two panels easy to compare by eye).
Superimposed to the shallow gradient, the colour distribution of galaxies in
the cluster line of sight (left panel) shows a clear excess at 
$B-V=0.98$, $V-R=0.62$ mag, marked with a circle in the figure.  

\begin{figure}
\psfig{figure=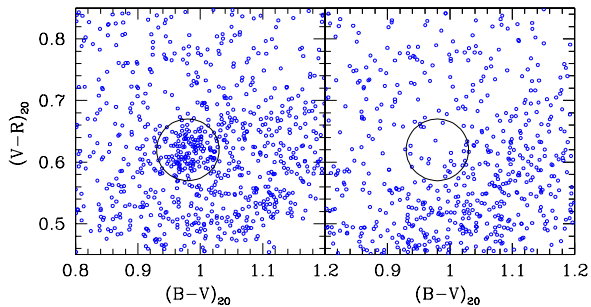,width=8truecm,clip=}
\caption[h]{
Colour-colour distribution of galaxies in the cluster (left panel) and
control field (right panel) lines of sight.  
An obvious overdensity is seen at
the colour of the red sequence ($B-V=0.98$, $V-R=0.62$ mag). Colours are
corrected to $R=20$ mag for the colour--magnitude effect. In the right
panel,
we randomly plot 1 out of 4.7 galaxies from the control field galaxies, in order 
to account for differences in surveyed areas.
} 
\end{figure}

\begin{figure}
\centerline{\psfig{figure=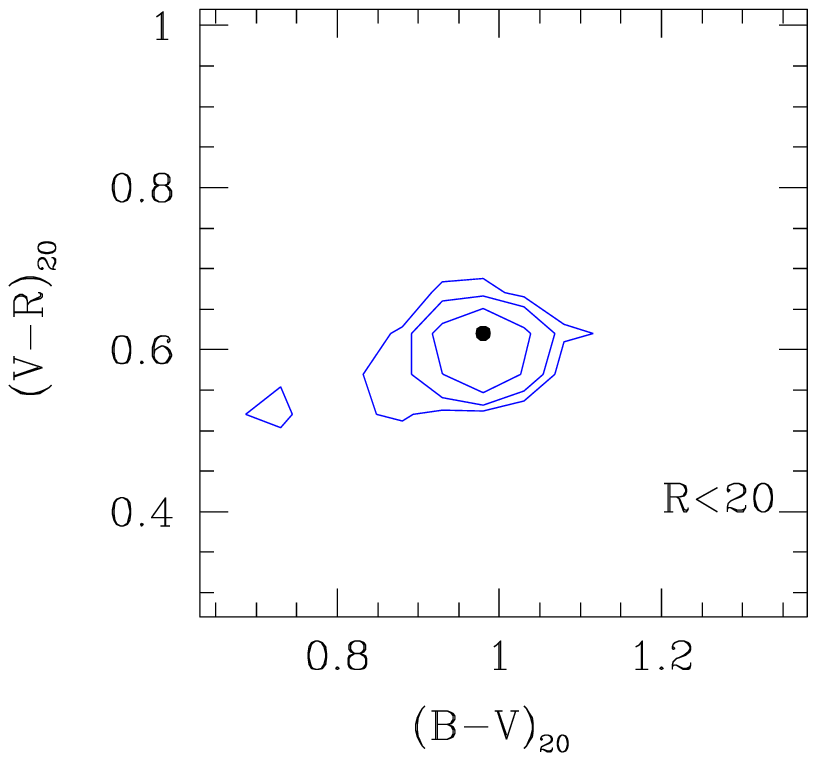,width=4truecm,clip=}%
\psfig{figure=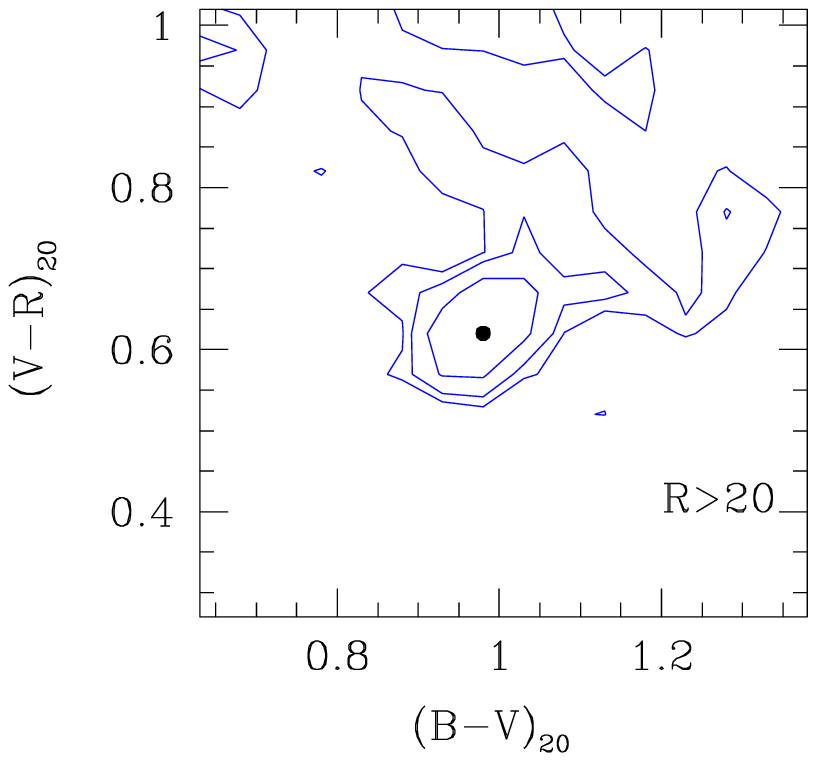,width=4truecm,clip=}}
\caption[h]{Background subtracted number density distribution of 
Abell\,1185 members in the 
colour--colour plane.
The left and right panels show the distribution of $R<20$ mag ($M_R=-15.7$ mag) 
and $R>20$ mag, respectively. Contours are draw to levels set at
1/2, 1/4 and 1/8 of the peak value. The extension in the right
panel toward red colours of the lowest contour has $S/N \sim 1$.
}
\end{figure}

Figure 3 shows the colour distribution, binned in colour bins and  
background subtracted using our control field.  The direction
that point to the observer is now the number density of Abell\,1185
members of a given colour. The two panels display the result for galaxies
brighter (left panel) and fainter (right panel) than $R=20$ mag
($M_R=-15.7$ mag). A clear peak at the colour of the colour-magnitude
relation is seen in both panels. The peaks are narrow: they have a width
of $\pm0.04$ mag (see appendix), implying that the thickness of the red sequence
is small. In Figure 3, the shown distributions are the observed
distributions convolved by a $0.1$ mag top-hat kernel. 
The two peaks are almost centered at the same colour ($B-V=0.98$, 
$V-R=0.62$ mag), shown as points in the figure. In the left panel there 
is an insignificant 0.02 mag offset in $V-R$, 1/5 of the width of the 
color bin.  This implies that the colour-magnitude relation does not deviate 
from linearity at faint magnitudes. A 0.1 mag shift would be very easily 
identified in the above plot. 

\begin{figure}
\psfig{figure=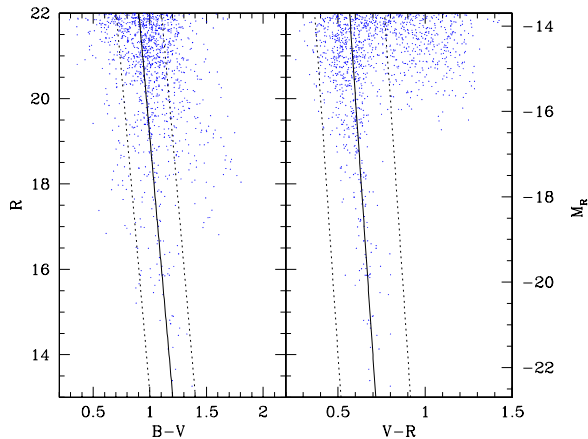,width=8truecm,clip=}
\caption[h]{
Colour--magnitude distribution of galaxies in the Abell\,1185 line of sight. 
In each panel are plotted galaxies within 0.2 mag from the red sequence in the other colour. 
Based on the control field images, we estimate that only one in five of
the plotted galaxies is a cluster member.}
\end{figure}

Figure 3 also shows that Abell\,1185 has a low fraction of blue galaxies 
in the observed portion (our images sample about half of the
cluster virial radius, which was computed from the cluster velocity
dispersion using equation 1 of Andreon et al. 2006): the peak does 
not show any large extension toward bluer (say, 0.2 mag or more) colours.
The S/N of the top--right part of lowest contours in right panel
of Fig 3 is $\sim 1$. Therefore, the
extension toward red colours of the $R>20$ distribution is hardly 
significant at all.

Figure 4 shows the distribution of galaxies in the colour--magnitude
plane of galaxies in the Abell\,1185 line of sight. 
Each panel shows galaxies falling in a plane of the colour-colour-magnitude cube of
thickness 0.4 mag in the other colour, in order to reduce the contamination by background galaxies. 
Because the spread around the red sequence is five times
smaller (see Figure 3 and Appendix), the 0.4 (i.e. $\pm0.2$) mag 
thickness is wide enough 
to avoid selectively removing any red sequence galaxy, and therefore
does not bias the slope of the red sequence. After the above selection,
we known from control field data that only one in five of the plotted
galaxies is a cluster member. The colour--magnitude relation in
$B-V$ and $V-R$ vs. $R$ are outstanding for galaxies brighter than $R=20$
or $21$ mag. At fainter magnitudes the background contribution makes 
the colour--magnitude relation difficult to disentangle. Figure 3
shows, however, that at $R>20$ the colour--magnitude relation does
not bend.

\begin{figure}
\psfig{figure=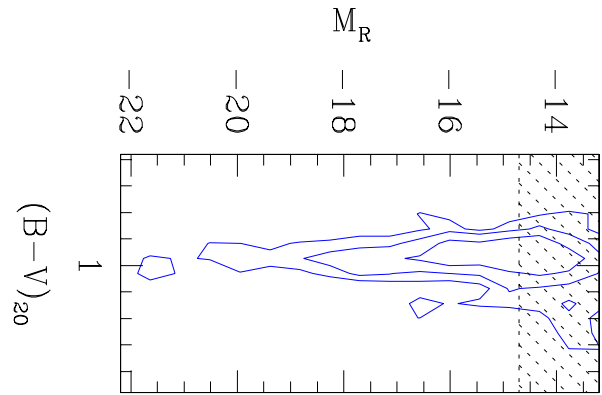,width=8truecm,clip=,angle=-90}
\caption[h]{
Background subtracted number density distribution of Abell\,1185 members 
in the colour--magnitude plane. Each panel shows
galaxies falling in a plane of the colour-colour-magnitude cube of
thickness 0.4 mag in the other colour, in order to reduce the 
contamination by background galaxies.}
\end{figure}

Figure 5 shows the number density distribution in the colour--magnitude
plane after background subtraction. In order to reduce the noise due to background fluctuations,
we considered only galaxies within $\pm0.1$ mag in $V-R$ from the red
sequence, which is a sensible choice given the small $V-R$ scatter  of red
galaxies. As in Figure 2 and 3, we have magnitude--corrected colours to
align the $R$ magnitude with one of the figure axis. As in Figure 3, the
data are convolved (binned) by a top-hat function of $\pm0.05$ mag
width in colour. The shaded region is affected by borders effects. 
The figure confirms the linearity of the red sequence and its
almost constant width.

The interpretation of these results is deferred to Sec. 4.4

We define as red galaxies the objects having colours
within 0.1 mag from both colour--magnitude relations. Such a definition
leaves intact the peak of the colour distribution centered on red galaxies
(see in particular Figure 3) and leaves room for objects scattered 
off from the red sequence by (our small) photometric colours. 

The sample of red galaxies turns out to be composed of about 180 members
plus a similar number of background galaxies. As a further check, we consider 
a second sample, relaxing the tightness of the colour constraint from $0.1$ to 
$0.2$ mag. The number of red galaxies remains unchanged, but the number of 
background galaxies grows by more than a factor four!

\begin{figure}
\psfig{figure=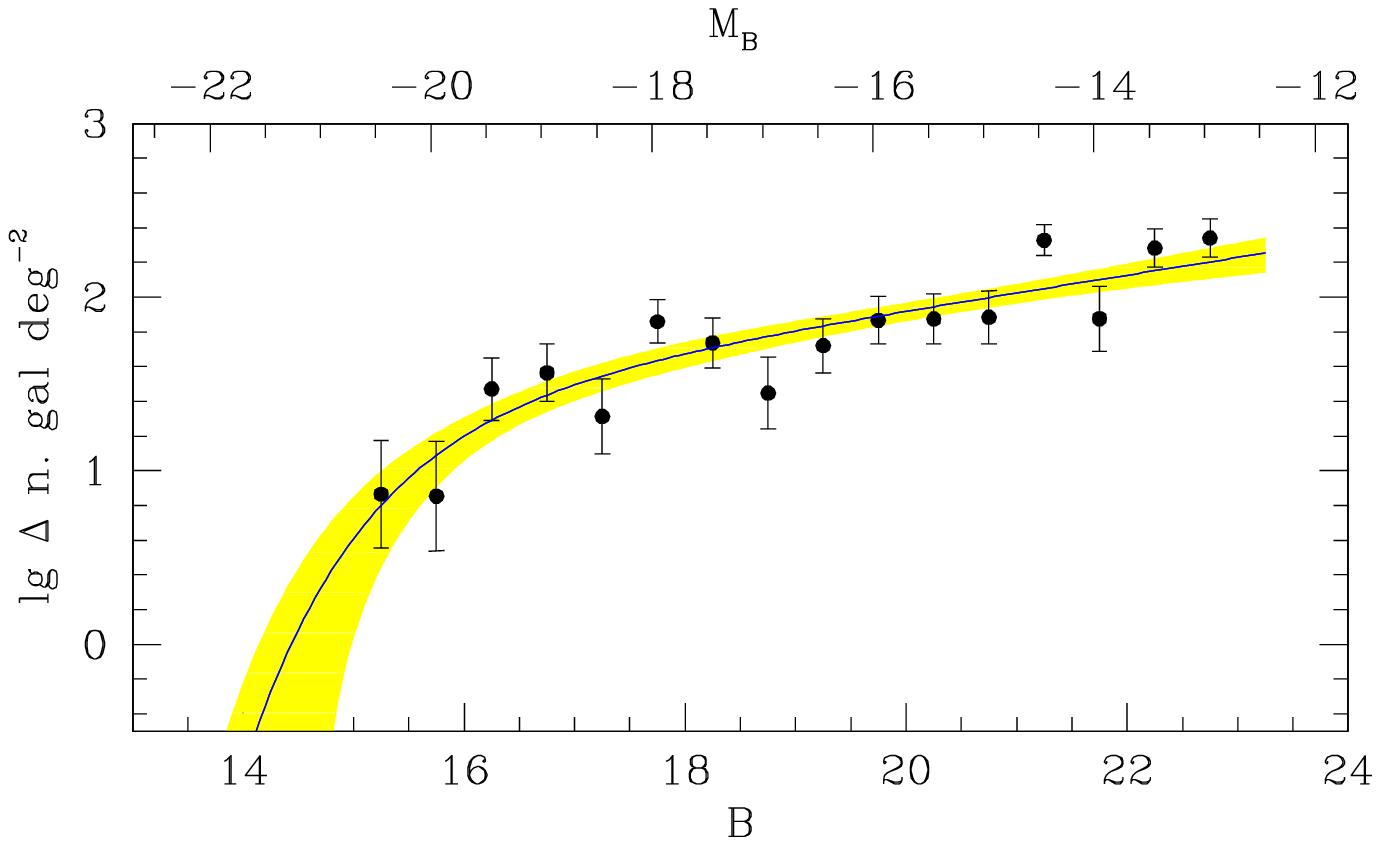,width=8truecm,clip=}
\psfig{figure=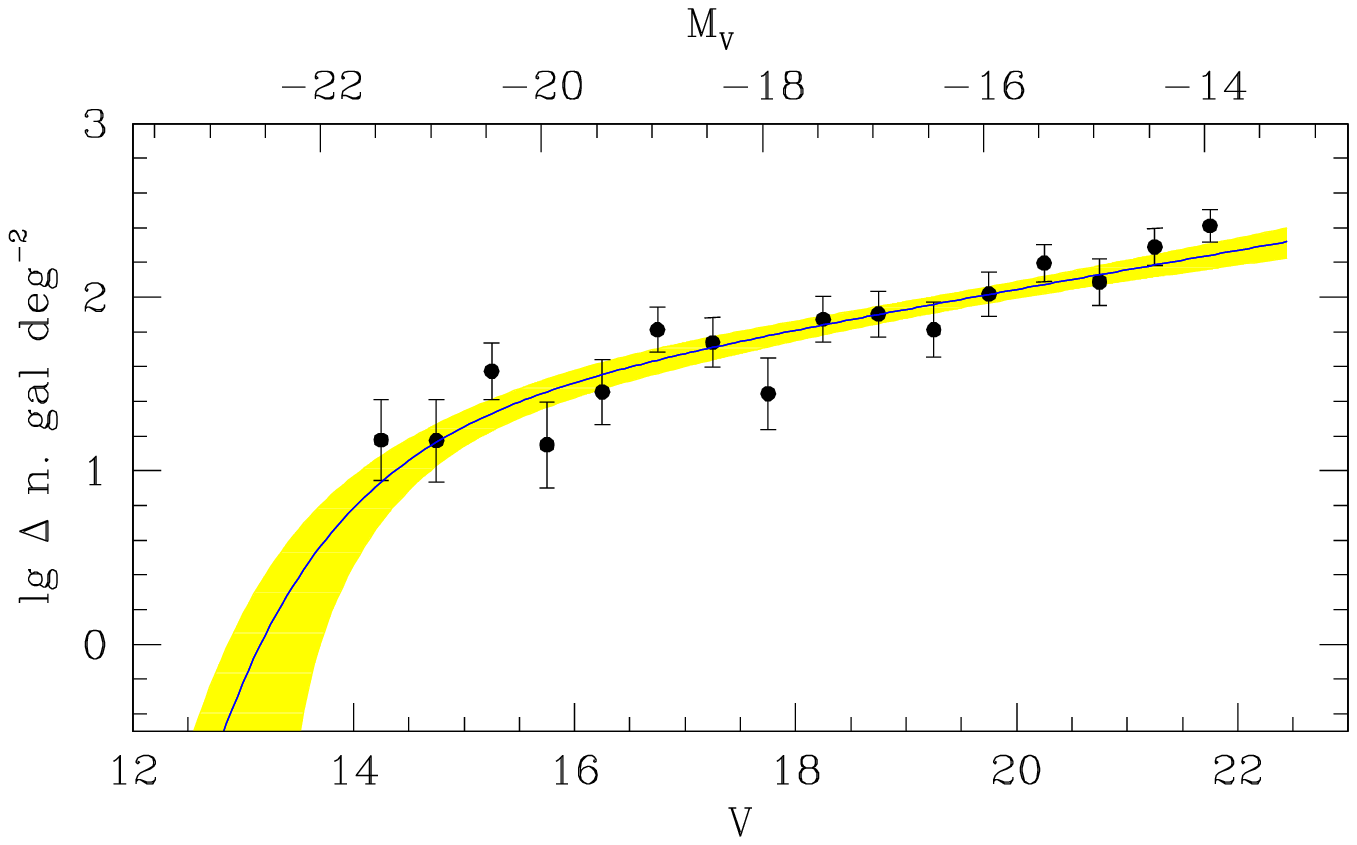,width=8truecm,clip=}
\psfig{figure=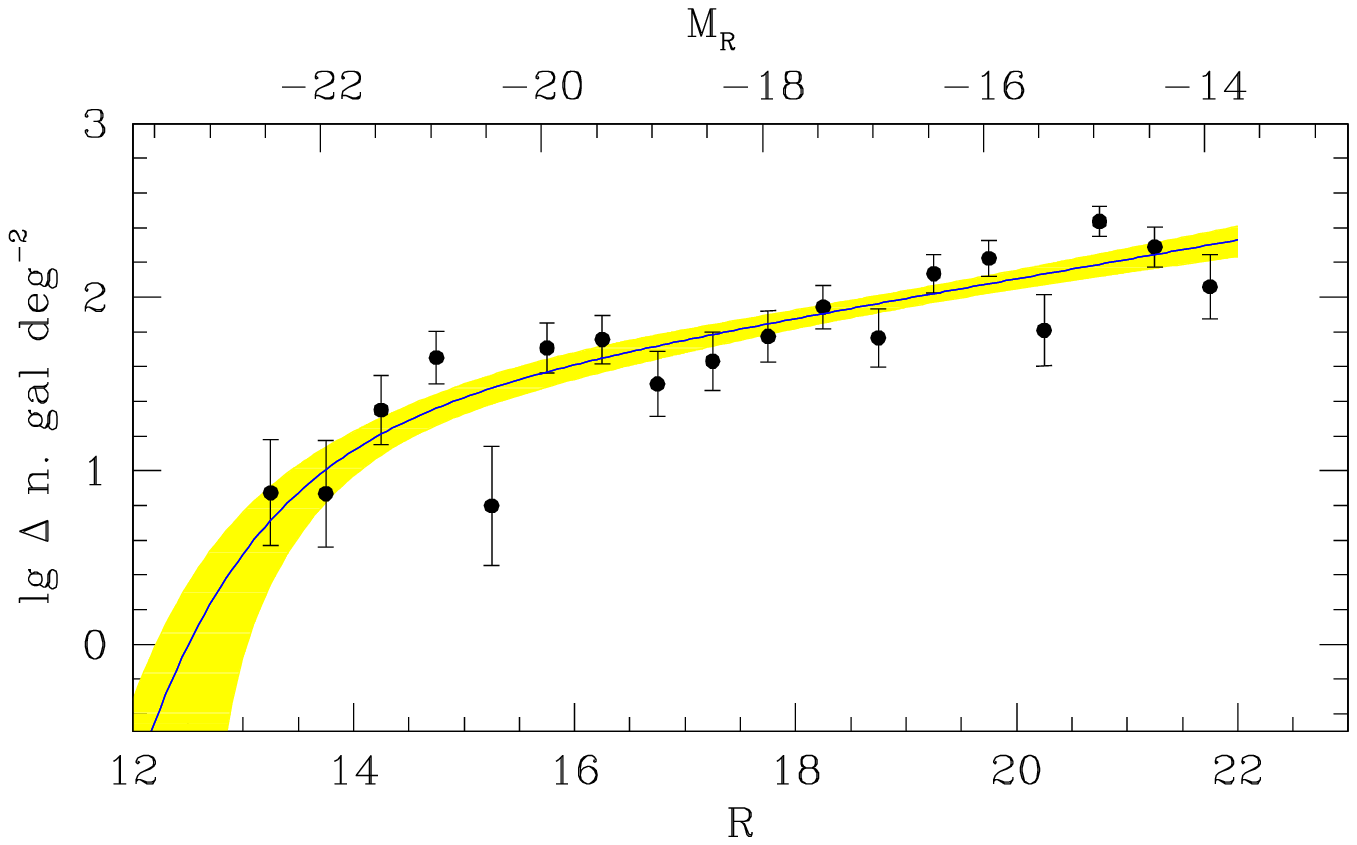,width=8truecm,clip=}
\caption[h]{$B$, $V$ and $R$ band LF of red galaxies in Abell\,1185.
Data points and error bars are derived by binning the data in magnitude
bins (and using an approximate error computation), whereas the curve
traces the best fit LF of unbinned data and the shaded area marks
the model uncertainty. 
} 
\end{figure}

\begin{figure}
\psfig{figure=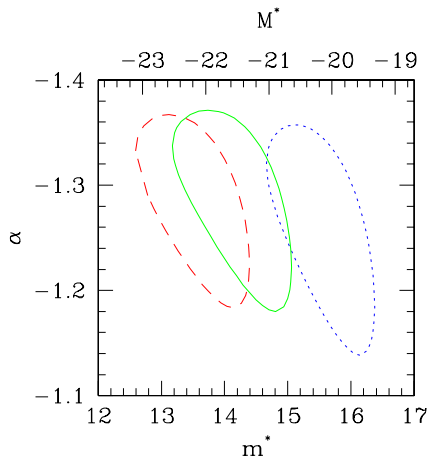,width=6truecm,clip=}
\caption[h]{$B$, $V$ and $R$ (from right to left)
68 \% confidence contours ($2\Delta \ln \mathcal{L} = 2.30$) 
on $m^*$ ($M^*$ in the top x-axis) and $\alpha$.  
} 
\end{figure}

\subsection{The Luminosity function}

\subsubsection{Methods}

Two methods are used to perform the statistical subtraction. For display
purposes, we use the traditional method, put forward by, say, Zwicky (1957)
and Oemler (1974) and summarized in many recent papers. The cluster
luminosity function is computed as the difference between galaxy counts in
the cluster and the control field directions, after binning the events (galaxy
magnitudes) in magnitude bins.  Approximate errors are computed as the
square root of the variance of the minuend, because the contribution due to
the uncertainty on the true value of background counts is negligible. 
Secondly, we fit the unbinned galaxy counts without any use of binned data
or errors computed in the previous approach. At this point we follow the rigorous
method set forth in APG, which is an extension of the Sandage, Tammann \&
Yahil method (1979, STY) to the case where a background population is present. 
The custom method adopts the extended likelihood instead of the conditional 
likelihood used by STY. The method (fully described in AGP) consists in fitting the unbinned distribution 
of clusters and control field counts using the likelihood, and in computing confidence
intervals using the likelihood ratio theorem (Wilks 1938, 1963). The advantage
of the rigorous method is that it provides unbiased results, reliable errors
estimations, hence assuring the correctness of the result.

\subsubsection{The LF of red galaxies}

Fig 6 shows the resulting LFs of red galaxies in the three filters: the binned data
with approximated error bars, and the best fit Schechter (1976) function with its 
uncertainty. Bins with incomplete data coverage are not plotted. The best fit parameters 
and their errors are listed in Table 1. This table also provides the range adopted for the 
fitting, a range such that the LF is unaffected by the $R<22.00$ mag initial selection. 

68 \% confidence contours are shown in Fig 7.

$m^*$ (or $M^*$) values, once values at the same slope $\alpha$ are 
considered, differ by the colour of red sequence galaxies (see Figure 7). 
$\alpha$ values in different bands are almost identical. This is expected,
because no passband dependency of the red galaxies LF is possible as long
as the colour-magnitude relation is linear. In fact, a linear mapping ($R$
to $B$ or $V$, in order to account for the colour-magnitude relation)
has a constant Jacobian and cannot alter the shape of a distribution function. 
The detection of a passband dependency for a sample selected to obey to a 
linear colour--magnitude relation is instead a sign of problems in 
the data analysis.

The slope of the LF is determined with good accuracy ($\pm0.06$) in all three 
bands but the uncertainty on the characteristic magnitude is large 
($\approx0.6$ mag), because we are studying a single not so rich cluster. 
We find that there are only about 20 red galaxies between $m_3$ and $m_3+2$ 
inside the studied portion of Abell\,1185.

The best fit parameters are robust to the adopted width of the colour
selection. Using a $\pm0.2$ mag width in place of our reference $\pm0.1$ mag 
width, we find in $R$: $m^*=13.9\pm0.5$ and $\alpha=-1.18\pm0.09$, in good 
agreement with the best fit parameters found by adopting a more stringent 
colour cut. Furthermore, 16 out of the 18 data points computed with the two 
colour selections differ each other by less than half the error bar in a 
non systematic way, clarifying that our reference colour boundary does not 
bias the LF. In particular, the above comparison shows that the Eddington 
bias is negligible, as expected.

\begin{table}
\caption{Fit results}
\begin{tabular}{lcll}
\hline
filter & range & $m^*$ & $\alpha$  \\
\hline
$B$ &	$B<23.25$ &	$15.7\pm 0.5$ & $-1.25\pm0.06$ \\
$V$ &	$V<22.45$ &	$14.3\pm 0.5$ & $-1.28\pm0.06$ \\
$R$ &	$R<22.00$ &	$13.7\pm 0.7$ & $-1.28\pm0.06$ \\
\hline
\end{tabular}
\hfill\break
\end{table}

\section{Comparison \& Discussion}

\subsection{Comparison with the red sequences from other clusters}

The red sequence of Abell\,1185 has none of the specific features that are
attributed to other clusters/environments by other works. The red sequence 
is linear and does not bend toward the blue at faint magnitudes, as
Conselice et al. (2002) found for candidate members of the Perseus cluster.
It neither bends toward the red as Evans et al. (1990) found for Fornax
galaxies. The small scatter at faint magnitudes is much smaller than
the half magnitude scatter that Conselice et al. (2002) found for candidate 
members of the Perseus cluster. Simply put in a few words, extremely faint
galaxies in Abell\,1185 follow the extrapolation of brighter red galaxies. 

We underline that each of these features we do not detect may be
reproduced by mismatches between cluster and control field images
or by an incorrect estimation of the galaxy membership. On the other
hand it is highly unlikely that these effects would render linear an 
intrinsically bended red sequence, or would strongly reduce the scatter 
of a scattered red sequence.

Our claim of a linear colour-magnitude of Abell\,1158 is in apparent 
disagreement with the Ferrarese et al. (2006) claim of a non-linear colour--magnitude 
in Virgo. In order to cross check their results, we take advantage of the fact 
that the authors generously shared their data. We consider two models: a 
linear and a quadratic colour--magnitude relation, as considered by the authors. 
We assume normal errors on colours and an intrinsic scatter, and we compute 
the likelihood following the laws of probability (the likelihood for this 
problem is presented in D'Agostini 2005 and in the appendix of Andreon 2006). 
Computation of evidence (e.g. 
Liddle et al. 2004, for an astronomical introduction) and information content 
(e.g. Trotta 2005) clarify that either data are non-informative or, if any, 
they support the simpler model (a linear CM), and surely do not support the 
extended model (i.e. the quadratic CM). We note that while Ferrarese et al. (2006) 
acknowledge the existence of an intrinsic scatter, they do not account for
it in the error computation (while it is accounted for in the present
statistical analysis).  Ignoring this component makes their fit to the 
bright and faint halves of their sample incompatible to each other, while 
they become compatible once the intrinsic scatter is accounted for.  

Our approach, instead, is informative because it differentiates between a
linear and non-linear model. Furthermore, we do not need any complex
statistical analysis to differentiate them because of our small errors. 
Any potential 0.05 mag CM drift away from the linear CM is equal to the 
total observed scatter (intrinsic plus photometric scatter) around the CM 
in Abell\,1185. For Virgo, it is one third of the observed scatter. 
In case of Abell\,1185, a 0.1 mag drift is easily spotted even by eye 
(for example in Figure 3), whereas in case of Virgo it requires an advanced 
statistical analysis that ought to conclude that the data are largely un-informative 
about the colour-magnitude curvature.

We emphasize that there is no colour selection in Figure 3, and therefore we do
not miss the possible CM bending (curvature) because of selection effects. A 0.05 mag 
drift away from a linear CM would lead to a distribution having different centers in 
the two panels by 0.05 mag, which is ruled out. In Figure 4, the colour selection 
($\pm0.2$ mag) is wider than a possible 0.05 mag drift and the CM does not become 
linear because its bending part is cut away by the colour selection.

\begin{figure}
\psfig{figure=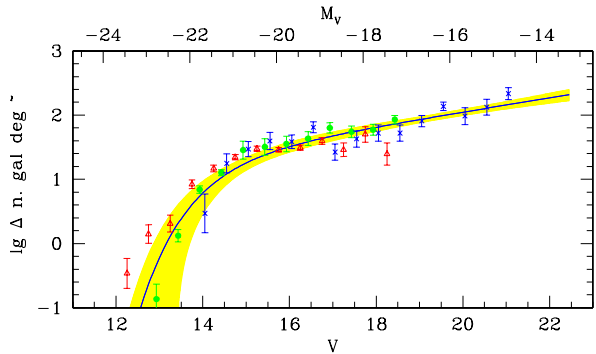,width=8truecm,clip=}
\caption[h]{$V$ band LF of red galaxies in Abell\,1185 (shaded region),
Coma (Secker \& Harris 1996; blue crosses), two $z=0.2$ clusters (Smail et al.
1998; green points), and nearby clusters in Tanaka et al. (2004; red 
triangles). All LFs are normalized (vertically shifted) to Abell\,1185 
down to $M_V=18$ mag. All LFs, but Tanaka et al. (2005), agree with each 
other. Abell\,1185 goes deeper than the other LFs and has smaller errors.} 
\end{figure}

\subsection{Comparison with literature red galaxies LF}

Figure 8 shows our $V$-band LF and its uncertainty (shaded region) and
some literature red galaxies LFs. All LFs have been normalized
to the richness (computed for galaxies brighter than $M_V=-18$ mag)
of Abell\,1185. Secker \& Harris (1996) compute the LF of
the Coma cluster in the $R$ band, converted to $V$ using the
observed $V-R$ colour of Abell\,1185 red galaxies. Smail et al.
(1998) compute the $I$ band LF of red galaxies in some clusters
at $z\sim0.25$, well matching $V$-band observations in the cluster
frame. In order to plot their data on Fig 8, we only need to
compute the (small, 0.1 mag) E+K correction, derived using
Bruzual \& Charlot (2003), and the distance modulus of the studied 
clusters. These authors provide two red LFs, one based on the use 
of an homogeneous cluster and control field, plotted in Figure 8, 
and a second one using heterogeneous data (which is not considered 
here for the present comparison, in light of the mentioned difficulties 
in using heterogeneous data for LF studies). The red galaxy LF of Abell\,1185
matches well these red galaxy LFs. Of course, every work adopts a slightly different 
definition of 'red galaxy', such as a using a different width around
the colour--magnitude (e.g. Secker \& Harris 1996 choice is twice
larger than ours), or increasing it with magnitude to account
for a larger photometric scatter at fainter magnitudes (e.g. Smail
et al. 1998), or using different filters (in both the observer and
cluster frames). The current agreement of these  LFs shows 
that the LF is robust to details in the 'red galaxy' definition. 

Figure 8 also shows the Tanaka et al. (2005) LF, computed
in V band like our LF. It is $\approx 0.75$ mag too bright, for reasons
that we are unable to understand.

Overall, our LF is in agreement with previous works. It is
deeper and has smaller errors than previously published LFs, and 
thanks to the rigorous analysis, it is considered more reliable.

\subsection{A more complex red galaxies LF?}

In this section we ask ourself if our LF model should be updated 
in regards of published LFs, i.e. if the red galaxy LF is cluster-dependent; 
in particular whether we should use a more complex LF in order to account 
for the possible existence of a dip, i.e. a region with depressed counts, 
a feature actively debated in the literature (e.g. Biviano et al. 1995; Secker \& Harris 1996).

We found no evidence for a dip in Abell\,1185 (see Figure 6) and therefore we looked
with further attention to the published claims about its existence. We consider with
special attention  Secker \& Harris (1996), because the authors excel in accounting 
for the uncertainty in the background subtraction and in providing all the
statistical details needed to re-analyse their data. They find that the LF of 
red galaxies in the Coma cluster is well modelled by a sum of a Gaussian and a 
Schechter, suggested by the presence of a possible dip in the Coma LF at $M_R\sim-19$ mag. 

The simple Schechter model that best fit Abell\,1185 data is a good description of 
their data (of Coma), being the $\chi^2=22$ for $15$ degree of freedom (i.e. larger $\chi^2$ 
are observed 10 \% of the times under the null hypothesis that the data are drawn from the
model), as can also qualitatively be seen in Fig 8. This comparison suggests
that, in light of our results on Abell\,1185 (which were of course not available at 
the time of their analysis) their data does not {\it require} a model more complex 
than a Schechter with $\alpha$ and $M^*$ fixed to the best fit Abell\,1185 values. 
The Bayesian Information Criteria (Schwartz 1978; Lindle 2004 provides a useful astronomical 
introduction) quantitatively informs us that evidence is largely in favor of the simplest model. 

We cannot repeat the above analysis to other published works because they
usually do not provide all the necessary details. Qualitatively, a model
with a dip seems unnecessary, even when the $p$ value of the data (computed 
using the $\chi^2$ or another statistic) is small (i.e. the fit 
with a Schecther function is claimed 
to be bad), since the increase in log-likelihood obtained by introducing the 
extra parameters is more than compensated by the huge increase in dimensionality 
of the parameter space over which likelihood should be averaged.

\subsection{Discussion on methods}

\subsubsection{Impact of the method used}

Compared to previously published works, the CM presented in this study has  been
determined in a fairly different way, and therefore it is important  to discuss
whether our results may depend or not on the analysis method. We  derived all
our results twice: first using the common method currently  used in literature,
and then using a statistical method based on axioms  of probabilities. The two
methods give the same results qualitatively but the second method has the
advantage of providing results with guaranteed correctness, at the difference of
a blind application of basic rules potentially giving misleading results, as
detailed also in APG, Andreon  et al. (2006), and also by several previous works
(e.g. Kraft et al. 1991;  Loredo 1992).

The most notable difference between our study and the previous works is 
that instead of subtracting the background of galaxies, we account for it 
more rigorously. The present work is a continuation of activities started 
in APG, and continued in Andreon et al. (2006) and Andreon (2006).

The usual astronomical recipe for background subtraction, $n_{clus}=n_{tot}-n_{bkg}$, 
called unbiased estimator by sampling theories (e.g. MacKay 2005), has a somewhat 
different meaning from what it seems. Let us suppose, for example, that 3 galaxies 
are observed in the cluster line of sight, and 5 are expected given our (assumed
to be quite precise) control field observations. This situation is possible, because 
of Poisson fluctuations. If we compute the cluster contribution as the difference 
between the above two numbers, we find a negative number of galaxies in the cluster
(or say in a given location of the parameter space, i.e. in a given range of colours 
and mag), leading to the unphysical result of non-positive numbers of galaxies.
Therefore, $n_{tot}-n_{bkg}$ is an estimator of $n_{clus}$ that in certain conditions 
has properties that are not acceptable.

A naive solution to avoid unphysical values is to set these negative values to
zero. However, this correction overestimates the total number of cluster
galaxies (i.e. the one of the whole sample), because negative fluctuations have
been corrected all in the same direction by increasing them, and positive
fluctuations left untouched (effectively a Malmquist-type bias). Another
possible naive solution is to widen the bin size (if data are binned), until
unphysical values disappear. Physically, it may happen that the bin becomes much
larger in a direction than the uncertainty of the data points it contain. In
such a situation, in order to remove an unphysical result we are forced first
to put in the same bin incompatible data (because at several $\sigma$'s away
from each other), and then to subtract off each other incompatible data, a 
procedure which can only lead to doubtful results. This situation often occurs 
at the less populated bright magnitudes, where errors are small.

Therefore, if $n_{tot}-n_{bkg}<0$, we are in an unsatisfactory position: we may 
choose between: a) having unphysical values at some location and an unbiased 
total number count, or b) having physical values for individual measurements 
and a biased value for the whole sample, or c) adding/subtracting each other 
incompatible data points. Actually, the same problem holds for $n_{tot}-n_{bkg}\ga 0$ 
(of which some aspects  are called Malmquist bias in literature, see Jeffreys 
1938 for a lucid discussion), and has already been discussed in Kraft et al. (1991), 
APG and Andreon et al. (2006) and others. As soon as the dimensionality of the 
data space is large, or when, as in the present case, the distribution of the 
data in the data space is uneven, there will always been some bins/location 
where $n_{tot}-n_{bkg}\approx 0$, i.e where some problems arise.

The second shortcoming of the usual astronomical recipe, $n_{tot}-n_{bkg}$, 
concerns the measurement of its uncertainty. When the difference above is negative,
a confidence interval (at whatever confidence level) is an empty interval
(Kraft et al. 1991). With the confidence interval empty, no interval can be
shorter, not even the one derived in absence of any background, an incoherent
result. For example, contrary to common sense, one may conclude because of
that approach that it makes no sense to perform a redshift survey to determine 
the individual membership of every galaxy in the sample, because these new 
observations will lead to larger confidence intervals! The null length of 
the confidence interval produces another absurd result: let us suppose 
that the derived confidence interval turns out to be of positive length. 
There is an easy way to make it smaller: by {\it adding} a background in 
the hope that Poisson fluctuations make $n_{tot}-n_{bkg}<0$. For a 
sufficiently large background the above occurs 50 \% of the times. 
Such a solution, although formally correct, is unacceptable intuitively
since it implies adding noise in order to reduce uncertainties! 

The third shortcoming of the usual astronomical recipe comes from binning 
the data. As discussed in APG, seldom the size, number and location of 
bins are considered, as well as their relevance for the final results. 
Does a trend has been missed because the non-optimal quantization of 
data in bins?

Bayesian methods do not present such troubles, because: a) they do not require 
that data are binned; b) they are squarely based on axioms of probabilities (i.e.
returning credible intervals in agreement with logic and common sense).
[START NEW TEXT]
Specifically, instead of removing the background, we introduced the
background component, and we then marginalize the posterior over the
nuisance (background) parameters, as required by the sum axiom of
probability. [END NEW TEXT]  c) 
they naturally account for boundaries in the data or parameter space.
In the present work several parameters have boundaries: for example, 
the intrinsic scatter of
the colour-magnitude relation and the number of cluster or field galaxies 
cannot be negative. 

Therefore, 
for display purposes we use astronomical recipes, which often hold, but adopted 
a full Bayesian computation for estimating parameters (and uncertainties) and 
for model selection.  
Appendix of Andreon (2006) describes in detail the stochastical 
computation using the Monte Carlo Marcov Chain (MCMC, Metropolis 1953) method used 
to determine the colour--magnitude relation.

The full Bayesian computation and the use of common rules, both provide compatible 
results: in fact, in all figures showing data points computed with astronomical recipes,
and models computed with probability axioms, the two derivations agree each other.
For example, curves and data points in Figure 6 agree with each other. The 
colour--magnitude slope derived from theory of probabilities (solid line in Figure 2) 
agrees data points in Figure 2. The full Bayesian computation is however not useless. 
The agreement between the two approaches has been ``guided'' by the more precise 
computation: for example, in Figure 6 by taking much smaller bins, we easily get 
negative numbers of cluster galaxies and also an unbelievable luminosity function 
that takes only discrete values (because cluster galaxies come in integer units),
an unphysical result. Furthermore, ``guided'' by the Bayesian analysis
in Figure 6 we did not plot confidence
intervals (that have the mentioned shortcoming  of not 
becoming small when the uncertainty on nuisance parameters
decreases and of potentially being
of null lenght), but we computed the (approximated) variance of
the estimator $n_{tot} -n_{bkg}$.
One of the advantages of Bayesian methods is that they 
{\it always} return numbers with guaranteed correctness, the disadvantage is that 
sometime they are expensive to implement computationally. 


The LF is determined strictly following APG, i.e. a maximum likelihood method,
to which we defer for details, instead of a determination from a fully Bayesian 
derivation.
In the present paper, we took advantage from the fact that regularity conditions for 
the use of the likelihood ratio theorem holds for LF computations, and therefore 
we used the inexpensive maximum likelihood method instead of cpu--expensive 
Bayesian approach for the stochastical computations. 

\subsubsection{The role of assumptions}

During our CM determination in the Appendix, we made several assumptions. First of all,
we determined the slope of the colour--magnitude assuming its linearity.
However if the colour--magnitude is bended, the meaning of ``slope" 
(a number, indeed) is ill defined: what is the slope of a bended line? 
Therefore, a valid question that precedes the slope determination is: 
does the colour--magnitude is bended or linear? This task is named model 
selection, and we met it when discussing this very same issue for Virgo 
galaxies (sec 4.1), and again when we commented about the existence of 
a dip in the LF (sec 4.2). As done in sec 4.1, 
to answer the question, it is just matter of
increasing the degree of the polynomial describing the colour--magnitude
relation shape and compare the relative evidence of the two models: 
the linear model turns out to be highly preferred. The outcome of this 
comparison is quite expected, the colour--magnitude is clearly linear,
as the inspection of Figure 4 at bright magnitudes clearly demonstrate, 
as well as Figure 3 at faint magnitudes, and Figure 5 over the whole 
magnitude range. The first two plots are two simple projections of the 
data cube, no analysis whatsoever has been applied. In Figure 5, instead, 
the background has been removed, by quantizing data in bins in the
colour--colour--magnitude plane. 

Second, we further assumed that the LF of red galaxies is described by a 
Schechter function and that the distribution of background galaxies can 
be modelled as a polynomial in the restricted range of colours we are 
interested in. The former assumption is justified by Figure 6 (see
the data points, derived without any assumption about the LF shape). 
The latter assumption is quite general: every distribution sufficiently 
smooth can be approximated by its Taylor expansion, which is a polynomial. 
Therefore, we raised the degree of the polynomial until a satisfactory 
match with the data is reached.

\subsubsection{Conclusions on methods}

The performed analysis squarely relies on laws of probabilities and accounts
for the existence of boundaries.
It is possible to follow a simpler path to derive the results, of course, 
but we cannot guarantee the correctness of the results obtained by
methods that ignore (or not fully account for) laws of probabilities and
ignore the existence of boundaries in the data or parameter space.


\subsection{Discussion on results}

The colour-magnitude relation  has a simple interpretation in the context
of galaxy formation: the brighter and more massive galaxies have deeper
gravitational potential wells, and therefore are more prone at retaining 
the interstellar gas which becomes superheated and metal enriched 
during initial stages of the star-formation epoch (Dekel \& Silk 1986).
Assuming that the magnitude blueing of the red sequence is only due to metallicity,
and adopting the Harris (1996) [Fe/H] vs $B-R$ calibration, galaxy metallicity
changes by 1.2 dex in the explored magnitude range. 
Although the colour-magnitude sequence is interpreted in terms
of a metallicity trend (Kodama \& Arimoto 1997), our data alone may
be interpreted in terms of age (fainter galaxies are younger and
thus bluer), because of the well known age/metallicity degeneracy 
(e.g. Worthey 1994).

Abell\,1185 does not present the appealing features claimed to have been 
observed in other clusters and discussed at length in previous published works. 
Its red sequence does not disappear at faint magnitudes, neither bends toward red or
blue colours: it simply continues straight, following the extrapolation of
what is usually observed at brighter magnitudes. The scatter along the CM does
not largely increase with magnitude as claimed for galaxies in the Perseus 
clusters. The LF of red galaxies shows no dip and no pass-band dependency.

All the above points offer us the chance of not looking for physical mechanisms
affecting the objects features in such a way to produce the (unobserved) heterogeneity.
The absence of features points toward an homogeneity of red galaxies over
the whole nine magnitude range explored (four decades in stellar mass).
Our argument is the usual one, and can be found on most papers
discussing the homogeneity of red galaxies (e.g. Bower, Lucey \& Ellis
1992; Andreon 2003), with the only difference that what is discussed for
bright galaxies holds here for the extreme faint galaxies: homogeneity 
in colour implies an old star age or a synchronization 
between the history of star formation of the various galaxies.

Faint ($M\approx M^*+3$ mag) red galaxies have attracted the 
attention of astronomers in recent days. At high redshift the colour 
magnitude relation sees signs of depletions, i.e. it has been reported 
that at $M^*+2.5$ the colour-magnitude relation disappears (Kodama et al. 2004) or
it is strongly underpopulated (De Lucia et al 2004). On the other
end, the effect does not appear to be universal, since the $z=0.83$ MS\,1054
cluster does not show any signs of depletion (Andreon, 2006) down
to $M^*+3.5$ mag.  We defer to Andreon (2006) for 
a detailed discussion about the reality of the claimed deficit
of faint red galaxies. 
Kodama et al. (2004) and De Lucia et al (2004)
conclude that faint ($M^*+3$) red galaxies are a recent ($z \la 1$) 
population. This interpretation does not fit our data: if faint red 
galaxies are younger than brighter red galaxies, then they should not 
lay on the extrapolation at faint magnitude of the CM relation observed 
at brighter magnitudes (i.e. the CM should bend toward the blue)
and the scatter should increase toward faint magnitudes (there
is less time for the objects to make their colour more similar).
Quantification of the offset to be observed is simple: if faint
galaxies form at $z\sim0.8$ instead of $z\sim3$, they should
be bluer by 0.06 mag in $B-V$ (assuming a single stellar population,
e.g. Buzzoni 2005), a drift that we do not see in Abell\,1185.

\section{Conclusions}

We have studied the colour of red galaxies of the Abell\,1185 cluster at
$z=0.0325$ down to $M^*+8$ in the $B$, $V$ and $R$ bands. The
colour--magnitude relation is linear without evidence for a significant
bending down to absolute magnitudes seldom probed in literature
($M_R=-12.5$ mag). It is also thin ($\pm0.04$ mag) and its thickness is not
a strong function of magnitude. The luminosity function of red galaxies in
Abell\,1185 is adequately described by a Schechter function, with
characteristic magnitude and faint end slope that also well describe the LF
of red  galaxies in other clusters. There is no passband dependency of the
LF shape,  other than an obvious $M^*$ shift due to the colour of the 
considered population. Red galaxies form an homogeneous population,
over four decades in stellar mass, for what concerns colours and
luminosity.
Homogeneity in colour implies an old star age or a synchronization 
between the history of star formation of the various galaxies, down to $M^*+8$.

\section*{Acknowledgments}

Stefano Andreon thanks Giancarlo Ghirlanda and Giovanni Punzi for useful discussion on the
68 \% confidence bounds, and Masayuki Tanaka for giving us their red LF in 
electronic form. Jean-Charles Cuillandre thanks Gregory Fahlman for giving
us access to discretionary time to realize this project.\\

Based on observations obtained at the Canada-France-Hawaii Telescope (CFHT) 
which is operated by the National Research Council of Canada, the Institut 
National des Sciences de l'Univers of the Centre National de la Recherche 
Scientifique of France, and the University of Hawaii.\\

Other facilities used (to which we defer for standard acknowledgements): NED.

\bsp

\appendix

\section{Thecnical details}

In section 3 we do not provide quantitative details about the way
the slope and scatter of the colour--magnitude relation of cluster
galaxies have been determined, a gap that we fill in this appendix.
We strictly follow Andreon (2006) who presents the statistical
approach, starting from axioms of probability and largely following
D'Agostini (2003; 2005). In short, we model the cluster distribution 
in the colour--magnitude plane as the product of 
a Schechter function in mag and a linear colour--magnitude relation 
with an intrinsic unknown scatter. Furthermore, we model 
the background distribution in the colour--magnitude plane as the product 
of a power law of degree two in mag and of degree one in colour times a 
Gaussian in colour with unknown dispersion whose central colour is 
linear depending on mag with unknown slope and intercept. 
Such a modeling requires seven nuisance parameters for the background, 
over which we  marginalize in order to derive the uncertainty on the interesting 
quantities (e.g. the parameters describing the colour-magnitude relation). 
The wide magnitude range and the aboundant background data considered in 
this paper, both oblige us to adopt a model with many parameters 
for the background and such a complicate model  
is still insufficient to describe the background 
distribution over the whole magnitude range accessible to the 
observations. Instead than increasing the model complexity of the
background (which encodes information which is not of our interest
in this paper), we prefers to 
discard the last magnitude bin where the background contribution is
so large than they carry a very poor information about
the slope and intercept of the colour--magnitude relation
(i.e. in our CM determination we use $R<21$ instead of $R<22$). 
One of the merit of this approach is not to claim plausible that
the CM has a negative scatter, feature that instead blesses
other methods that claim that negative (and therefore unphysical) values 
of the scatter are within the 68 \% confidence interval.

As in Andreon (2006), we are not interested in modelling the distribution of
red galaxies at colours where they are not observed and therefore we limit
the analysis to the $0.4<V-R<0.9$ mag and $0.7<B-V<1.5$ mag ranges.
Assuming uniform priors zero-ed in the unphysical ranges, we found: 

$B-V=0.986 \pm 0.007 -(0.038\pm0.003) (R-19.556)$ mag with 
a scatter of 0.039 mag and 

$V-R=0.608\pm0.007 -(0.016\pm0.002) (R-19.556)$ mag with 
a scatter of 0.036 mag.

\label{lastpage}


\begin{thebibliography}{}

\bibitem[Abell(1958)]{1958ApJS....3..211A} 
Abell, G.~O.\ 1958, ApJS, 3, 211 

\bibitem[Andreon(2003)]{2003A&A...409...37A} 
Andreon, S.\ 2003, A\&A, 409, 37 

\bibitem[Andreon(2006)]{} 
Andreon, S.\ 2006, MNRAS, 369, 969
 
\bibitem[Andreon \& Cuillandre(2002)]{2002ApJ...569..144A} 
Andreon, S., \& Cuillandre, J.-C.\ 2002, ApJ, 569, 144 

\bibitem[Andreon et al.(2004)]{2004MNRAS.349..889A} 
Andreon, S., Lobo, C., \& Iovino, A.\ 2004, MNRAS, 349, 889 

\bibitem[Andreon et al.(2005)]{2005MNRAS.360..727A} 
Andreon, S., Punzi, G., \& Grado, A.\ 2005, MNRAS, 360, 727 

\bibitem[Andreon et al.(2006)]{2006MNRAS.365..915A} 
Andreon, S., Quintana, 
H., Tajer, M., Galaz, G., \& Surdej, J.\ 2006, MNRAS, 365, 915 

\bibitem[Baldry et al.(2004)]{2004ApJ...600..681B} 
Baldry, I.~K., 
Glazebrook, K., Brinkmann, J., Ivezi{\'c}, {\v Z}., Lupton, R.~H., Nichol, 
R.~C., \& Szalay, A.~S.\ 2004, ApJ, 600, 681 
 
\bibitem[Bertin \& Arnouts 1996]{1996A&AS..117..393B} 
Bertin, E. \& Arnouts, S. 1996, A\&AS, 117, 393 

\bibitem[Biviano et al.(1995)]{1995A&A...297..610B} 
Biviano, A., Durret, 
F., Gerbal, D., Le Fevre, O., Lobo, C., Mazure, A., \& Slezak, E.\ 1995, 
A\&A, 297, 610 
 
\bibitem[Blanton et al.(2005)]{2005ApJ...631..208B} 
Blanton, M.~R., Lupton, 
R.~H., Schlegel, D.~J., Strauss, M.~A., Brinkmann, J., Fukugita, M., \& 
Loveday, J.\ 2005, ApJ, 631, 208 

\bibitem[Bower et al.(1992)]{1992MNRAS.254..601B} 
Bower, R.~G., Lucey, J.~R., \& Ellis, R.~S.\ 1992, MNRAS, 254, 601 

\bibitem[Bruzual A.~\& Charlot(1993)]{1993ApJ...405..538B} 
Bruzual A., G., \& Charlot, S.\ 1993, ApJ, 405, 538 

\bibitem[Conselice et al.(2002)]{2002AJ....123.2246C} 
Conselice, C.~J., Gallagher, J.~S., \& Wyse, R.~F.~G.\ 2002, AJ, 123, 2246 
 
\bibitem[Cuillandre et al. (2000)]{}
Cuillandre, J.-C., Luppino, G., Starr, B. \& Isani, S., 2000, SPIE, 4008, 1010

\bibitem[D'Agostini(2003)]{}
D'Agostini, G.\ 2003, "Bayesian reasoning in data analysis - A critical introduction", 
World Scientific Publishing

\bibitem[D'Agostini(2005)]{2005physics..11182D} 
D'Agostini, G.\ 2005, preprint, (physics/0511182) 

\bibitem[Dekel \& Silk(1986)]{1986ApJ...303...39D} 
Dekel, A., \& Silk, J.\ 1986, ApJ, 303, 39 

\bibitem[De Lucia et al.(2004)]{2004ApJ...610L..77D} 
De Lucia, G., et al.\ 2004, ApJ, 610, L77 

\bibitem[Drinkwater et al.(1999)]{1999ApJ...511L..97D} 
Drinkwater, M.~J., Phillipps, S., Gregg, M.~D., Parker, Q.~A., Smith, R.~M., Davies, J.~I., 
	Jones, J.~B., \& Sadler, E.~M.\ 1999, ApJ, 511, L97 

\bibitem[Evans et al.(1990)]{1990MNRAS.245..164E} 
Evans, R., Davies, J.~I., \& Phillipps, S.\ 1990, MNRAS, 245, 164 

\bibitem[Ferrarese et al.(2006)]{2006astro.ph..2297F} 
Ferrarese, L., et  al.\ 2006, ApJS, in press (astro-ph/0602297) 

\bibitem[Garilli Maccagni \& Andreon 1999]{1999A&A...342..408G} 
Garilli, B., Maccagni, D. \& Andreon, S. 1999, A\&A, 342, 408 

\bibitem[Jeffreys(1938)]{1938MNRAS..98..190J} 
Jeffreys, H.\ 1938, MNRAS, 98, 190 

\bibitem[Jones \& Forman(1984)]{1984ApJ...276...38J} 
Jones, C., \& Forman, W.\ 1984, ApJ, 276, 38 

\bibitem[Harris(1996)]{1996AJ....112.1487H} 
Harris, W.~E.\ 1996, AJ, 112, 1487 

\bibitem[Kodama et al.(2004)]{2004MNRAS.350.1005K} 
Kodama, T., et al.\ 2004, MNRAS, 350, 1005 

\bibitem[Kodama \& Arimoto(1997)]{1997A&A...320...41K} 
Kodama, T., \&  Arimoto, N.\ 1997, A\&A, 320, 41 

\bibitem[Landolt 1992]{1992AJ....104..340L} 
Landolt, A. U. 1992, AJ, 104, 340 

\bibitem[Liddle(2004)]{2004MNRAS.351L..49L} 
Liddle, A.~R.\ 2004, MNRAS, 351, L49 

\bibitem[]{}
Loredo, T., 1992, in "Statistical Challenges in Modern Astronomy", 
eds. E. D. Feigelson and G. J. Babu (New York: Springer-Verlag) 
pag. 275

\bibitem[]{} 
MacKay D., 2005, Information theory, Inference and Learning Algorithms,
Cambridge University Press

\bibitem[Magnier \& Cuillandre(2004)]{2004PASP..116..449M} 
Magnier, E.~A., \& Cuillandre, J.-C.\ 2004, PASP, 116, 449 

\bibitem[Mahdavi et al.(1996)]{1996AJ....111...64M} 
Mahdavi, A., Geller,  M.~J., Fabricant, D.~G., Kurtz, M.~J., 
	Postman, M., \& McLean, B.\ 1996,  AJ, 111, 64 

\bibitem[Mateo(1998)]{1998ARA&A..36..435M} 
Mateo, M.~L.\ 1998, ARA\&A, 36, 435 

\bibitem[McCracken et al.(2003)]{2003A&A...410...17M} 
McCracken, H.~J., et al.\ 2003, A\&A, 410, 17 

\bibitem[Oemler(1974)]{1974ApJ...194....1O} 
Oemler, A.~J.\ 1974, ApJ, 194, 1 

\bibitem[Sandage, Tammann, \& Yahil(1979)]{1979ApJ...232..352S} 
Sandage,  A., Tammann, G.~A., \& Yahil, A.\ 1979, ApJ, 232, 352 

\bibitem[Secker et al.(1997)]{1997PASP..109.1377S} 
Secker, J., Harris, W.~E., \& Plummer, J.~D.\ 1997, PASP, 109, 1377 
 
\bibitem[Secker \& Harris(1996)]{1996ApJ...469..623S} 
Secker, J., \& Harris, W.~E.\ 1996, ApJ, 469, 623 

\bibitem[Schechter(1976)]{1976ApJ...203..297S} 
Schechter, P.\ 1976, ApJ, 203, 297 

\bibitem[]{}
Schwarz G., 1978, Annals of Statistics, 5, 461

\bibitem[Smail et al.(1998)]{1998MNRAS.293..124S} 
Smail, I., Edge, A.~C.,  Ellis, R.~S., \& Blandford, R.~D.\ 1998, MNRAS, 293, 124 
 
\bibitem[Tanaka et al.(2005)]{2005MNRAS.362..268T} 
Tanaka, M., Kodama, T., 
Arimoto, N., Okamura, S., Umetsu, K., Shimasaku, K., Tanaka, I., \& Yamada, 
T.\ 2005, MNRAS, 362, 268 

\bibitem[Trotta(2005)]{2005astro.ph..4022T} Trotta, R.\ 2005, preprint, 
(astro-ph/0504022) 
 
\bibitem[Visvanathan \& Sandage(1977)]{1977ApJ...216..214V} 
Visvanathan,  N., \& Sandage, A.\ 1977, ApJ, 216, 214 
 
\bibitem[]{} 
Wilks, S., 1938, Ann. Math. Stat. 9, 60 

\bibitem[]{} 
Wilks, S., 1963, Mathematical Statistics (Princeton: Princeton University
Press). 

\bibitem[Worthey(1994)]{1994ApJS...95..107W} 
Worthey, G.\ 1994, ApJS, 95, 107 

\bibitem[Zwicky 1957]{1957moas.book.....Z} 
Zwicky, F.  1957,  Morphological Astronomy, Berlin: Springer

\end{thebibliography}
\end{document}